\documentclass{article}
\usepackage[english]{babel}
\usepackage{graphicx,subfigure}
\usepackage{amsmath}
\usepackage{subeqnarray}
\usepackage{amsgen}[1995/01/01]
\usepackage{amsfonts}[1995/01/01]
\usepackage{amssymb}[1995/01/01]
\usepackage{amsbsy}[1995/01/01]
\begin{document}
\newcommand{\keywords}[1]{\par\addvspace\baselineskip
\noindent\keywordname\enspace\ignorespaces#1}
\newcommand{\ee}{\`e\ }
\renewcommand{\aa}{\`a\ }
\newcommand{\oo}{\`o\ }
\newcommand{\uu}{\`u\ }
\newcommand{\ket}[1]{\ensuremath {|\: #1 \: \rangle}}
\newcommand{\bra}[1]{\ensuremath{\langle \: #1 \:|}}
\newcommand{\braket}[2]{\ensuremath{\langle \: #1 \: | \: #2 \: \rangle}}
\newcommand{\ketbra}[2]{\ensuremath{| \: #1 \:\rangle \langle \: #2 \:  |}}
\newcommand{\ves}[2]{\ensuremath{#1_1,#1_2, \ldots, #1_{#2}}}
\newcommand{\ve}[2]{\ensuremath{#1(1),#1(2) \ldots, #1(#2)}}
\newcommand{\Hc}{\ensuremath{\mathcal{H}_{cursor\;}}}
\newcommand{\Hr}{\ensuremath{\mathcal{H}_{register\;}}}
\newcommand{\Hcr}{\ensuremath{\mathcal{K}_{cur\;}}}
\newcommand{\Hrr}{\ensuremath{\mathcal{K}_{reg\;}}}
\newcommand{\Hm}{\ensuremath{\mathcal{H}_{machine\;}}}
\newcommand{\eref}[1]{(\ref{#1})}
%
\title {Quantum walks: a Markovian perspective}

\author{
\small{Diego de Falco and Dario Tamascelli}\\
\small{Dipartimento di Scienze dell'Informazione}\\
\small{Universit\`a degli Studi di Milano}\\
\small{via Comelico 39, 20135 Milano, Italy}\\
\small{\itshape{e-mail: defalco@dsi.unimi.it, tamascelli@dsi.unimi.it}}
\and
\small{CIMAINA, Centro Interdipartimentale Materiali e Interfacce Nanostrutturati,}\\
\small{Universit\`a degli Studi di Milano}
}
\date{}
\maketitle%

\maketitle
\begin{abstract}
For a \emph{continuous-time} quantum walk on a line the  variance  of the position observable grows quadratically in time, whereas, for its classical counterpart on the same graph, it exhibits a linear, diffusive, behaviour. A \emph{quantum} walk, thus, propagates at a rate which is linear in time,  as compared to the square root rate for a classical \emph{random} walk. Indeed, it has been suggested that there are graphs that can be traversed by a \emph{quantum} walker exponentially faster than  by the classical \emph{random} analogue. In this note we adopt the approach of exploring the conditions to impose on a Markov process in order to emulate its quantum counterpart: the central issue that emerges  is the problem of taking into account, in the numerical generation of each sample path, the \emph{causative} effect  of the ensemble of trajectories to which it belongs. How to deal numerically with this problem is shown in a  paradigmatic example.\\
\end{abstract}
\section{Paradigmatic examples} \label{sec:paradigmatic}
The identity 
\begin{equation}
  \sum_{x=- \infty}^{+ \infty} J_x(t)^2 =1,
\end{equation}
satisfied by the Bessel functions of first kind and integer order $J_x(t)$, shows that the function
\begin{equation}
 \rho(t,x)=J_x(t)^2
\end{equation}
is, for each time $t$, a probability mass function on the relative integers.\\[5pt]
We raise here the question of finding examples of \emph{phenomena} of probabilistic time evolution described by this probability mass function.\\[5pt]
We will give two \emph{distinct}, apparently very different, answers to the above question: finding the relationship between the two distinct examples we are going to exhibit below and discussing the extent and generality of this relationship will be the main focus of this paper.
\subsection{First Example}
\begin{figure}[t] \label{fig:figura1}
\centering
 \includegraphics[width=13cm]{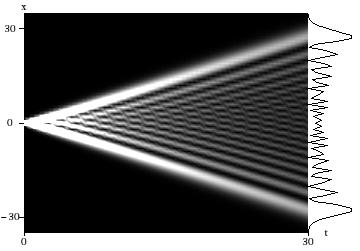}
\caption{A density plot of $\rho(t,x)=|\psi(t,x)|^2$. The profile of $\rho(30,x)$ as a function of $x$ is shown on the right.}
\end{figure}

The function
\begin{equation} \label{eq:solschro}
 \psi(t,x)=i^x J_x(t)
\end{equation}
is the solution of the Schr\"odinger equation on the relative integers
\begin{equation} \label{eq:discrschro}
 i \frac{d}{dt} \psi(t,x) =-\frac{1}{2} (\psi(t,x-1)+\psi(t,x+1))
\end{equation}
under the initial condition
\begin{equation}
 \psi(0,x)=\delta_{0,x}.
\end{equation}
Otherwise stated, the function $\rho(t,x)=J_x(t)^2$ is, at every time $t$, the probability distribution of a \emph{continuous-time} quantum walk on the graph having the relative integers as vertices, with edges between nearest neighbour sites \cite{childs02}. This quantum walk starts at time $0$ from the origin.\\
Figure \ref{fig:figura1}, a density plot of $\rho(t,x)=J_x(t)^2$, clearly shows the linear propagation expected in such a quantum walk. The reader more intersted in the \emph{phenomenon} than in the equation (in this case eq. \eref{eq:discrschro}) will appreciate recognizing in figure \ref{fig:figura1} the intensity pattern of propagation of light in a waveguide lattice \cite{perets07}.
\subsection{Second example}
Consider a birth-and-death random process $q(t)$ on the relative integers, evolving according to the following rules:
\begin{enumerate}
\def\theenumi{\roman{enumi}}
\item \label{it:geometric} \emph{geometric mean rule}: for every edge $\{x,x+1\}$ and every time $t$ the fraction of transitions per unit time taking place along this edge (number of transitions $x \to x+1$ \emph{plus} number of transitions $x+1 \to x$ per unit time)/(sample size) is equal to the geometric mean of the probability of the process being in $x$ and the probability of being in $x+1$;
\item \label{it:flux} \emph{local unidirectionality rule}: for every edge $\{x,x+1\}$, and depending on time $t$, only transitions $x \to x+1$ or only transitions $x+1 \to x$ are allowed;
\item \label{it:attractor} \emph{``horror vacui'' rule}: for every site $x$, if at a time $t_x$ the probability of being in $x$ passes through the value $0$, then there is an interval of time following $t_x$ in which along the edges $\{x-1,x\}$ and $\{x,x+1\}$ only transitions toward $x$ are allowed; this time interval terminates as soon as the probability of being in one of the two neighbours of $x$ crosses the value $0$ (at which instant the \emph{``horror vacui'' rule} takes hold for such a neighbour).
\end{enumerate}
As to the initial conditions, we suppose that there exists $\tau_0>0$ such that, for every integer $x$, 
\begin{equation}
 \rho(t,x) \equiv P(q(t)=x)>0,\mbox{ for $0<t<\tau_0$}
\end{equation}
and
\begin{equation}
 \lim_{t \to 0^+} \rho(t,x) =\delta_{0,x}.
\end{equation}
Together with the above initial condition on the position of the process, we impose, as a condition on its initial ``velocity'', the requirement that in the time interval $[0,\tau_0)$ only transitions taking the process \emph{away} from the origin are allowed.\\
We, finally, impose a left-right symmetry on the position of the process, in the form
\begin{equation}
 \rho(t,x)=\rho(t,-x)
\end{equation}
and a left-right symmetry on its ``velocity'' expressed in terms of its birth rate $\lambda(t,x)$ and its death rate $\mu(t,x)$ as
\begin{equation}
 \lambda(t,x)=\mu(t,-x).
\end{equation}
The transition probabilities per unit time $\lambda(t,x)$ and $\mu(t,x)$ are defined, respectively, by
\begin{equation} \label{eq:semlambda}
 p(t+\tau,x+1;t,x)=\tau \cdot \lambda(t,x)+o(\tau)
\end{equation}
\begin{equation} \label{eq:semmu}
 p(t+\tau,x-1;t,x)=\tau \cdot \mu(t,x)+o(\tau)
\end{equation}
for $\tau \to 0^+$.\\
Here and elsewhere we indicate by $p(t,x;t_0,x_0)$ the conditional probability
\[
P(q(t)=x|q(t_0)=x_0)
\]
of finding the process at time $t$ in $x$, given that at time $t_0$ it is in $x_0$.\\
Condition \eref{it:geometric} can, now, be written as the equation
\begin{equation} \label{eq:geometricmean}
 \lambda(t,x) \rho(t,x)+\mu(t,x+1)\rho(t,x+1)=\sqrt{\rho(t,x)\rho(t,x+1)},
\end{equation}
relating the three unknown fields $\lambda$, $\mu$ and $\rho$. The left hand side is, indeed, the probability per unit time of a transition along the link $\{x,x+1\}$. Notice that, because of \eref{it:flux}, equation \eref{eq:geometricmean} allows, locally, to express $\lambda$ or $\mu$ as a function of the values of $\rho$ at two neighbouring points.\\
A further equation involving the unknown fields is the continuity equation
\begin{eqnarray} \label{eq:continuity}
 \frac{d}{dt}\rho(t,x) & = & (\mu(t,x+1) \rho(t,x+1)-\lambda(t,x) \rho(t,x)) + \\
& + & (\lambda(t,x-1)\rho(t,x-1)-\mu(t,x)\rho(t,x)), \nonumber
\end{eqnarray}
expressing the fact that the probability mass at $x$ increases because of transitions $x \pm 1 \to x$ and decreases because of transitions $x \to x \pm 1$.\\
In the time interval $[0,\tau_0)$, we can therefore write, using also the left-right symmetry and the initial condition of allowing only transitions taking the process away from the origin (namely, for $0 \leq t <\tau_0$, $\lambda(t,x)>0$ for $x \geq 0$ and $\mu(t,x)>0$ for $x \leq 0$),
\begin{subeqnarray} \label{eq:continuitytau1}
\frac{d}{dt}\rho(t,0)& = &-2\sqrt{\rho(t,0)\rho(t,1)} \slabel{eq:continuitytau11}\\
\frac{d}{dt}\rho(t,x)& = &+\sqrt{\rho(t,x-1)\rho(t,x)}-\sqrt{\rho(t,x)\rho(t,x+1)},\mbox{ for $x > 0$}.  \slabel{eq:continuitytau12} \label{eq:continuitytau13}
\end{subeqnarray}
Equations \eref{eq:continuitytau1} are satisfied by $\rho(t,x)=J_x(t)^2$, for values of $t$ such that $J_x(t)$ is positive for every non negative integer $x$. This determines the numerical value of $\tau_0$ to be the smallest positive solution of the equation $J_0(t)=0$, namely
\begin{equation}
 \tau_0=2.4048.
\end{equation}
For a suitable value of $\tau_1>\tau_0$ condition \eref{it:attractor} will allow, in the time interval $[\tau_0,\tau_1)$, for transitions $\pm 1 \to 0$, so that equations \eref{eq:continuitytau1} are to be substituted, in this interval, by
\begin{subeqnarray} \label{eq:continuitytau2}
 \frac{d}{dt}\rho(t,0)& = & +2\sqrt{\rho(t,0)\rho(t,1)}, \\
 \frac{d}{dt}\rho(t,1) & =& - \sqrt{\rho(t,0)\rho(t,1)} - \sqrt{\rho(t,2)\rho(t,1)}\\
 \frac{d}{dt}\rho(t,x)& = &+\sqrt{\rho(t,x-1)\rho(t,x)}-\sqrt{\rho(t,x)\rho(t,x+1)},\mbox{ for $x > 1$}.
\end{subeqnarray}
Equations \eref{eq:continuitytau2} are again satisfied by $\rho(t,x)=J_x(t)^2$, but, this time, for values of $t$ such that $J_0(t)<0$ and $J_x(t)$ is positive for every positive integer $x$. This determines the numerical value of $\tau_1$ to be the smallest positive root of $J_1(t)=0$, namely
\begin{equation}
\tau_1=3.8317.
\end{equation}
The above considerations can be iterated: using the fact that between two consecutive zeroes of $J_x(t)$ there is one and only one zero of $J_{x+1}(t)$, one can control the changes of sign determined by \eref{it:attractor} in the continuity equation, to the effect of proving that the process $q(t)$ described by the conditions posed above satisfies, for every $t$, the condition
\begin{equation} \label{eq:soluzione}
 \rho(t,x) \equiv P(q(t)=x)=J_x(t)^2.
\end{equation}
Figure 2, to be compared with figure 1, shows a few sample paths of the process $q(t)$.
\begin{figure}[t] \label{fig:samplepaths}
\centering
 \includegraphics[width=12cm]{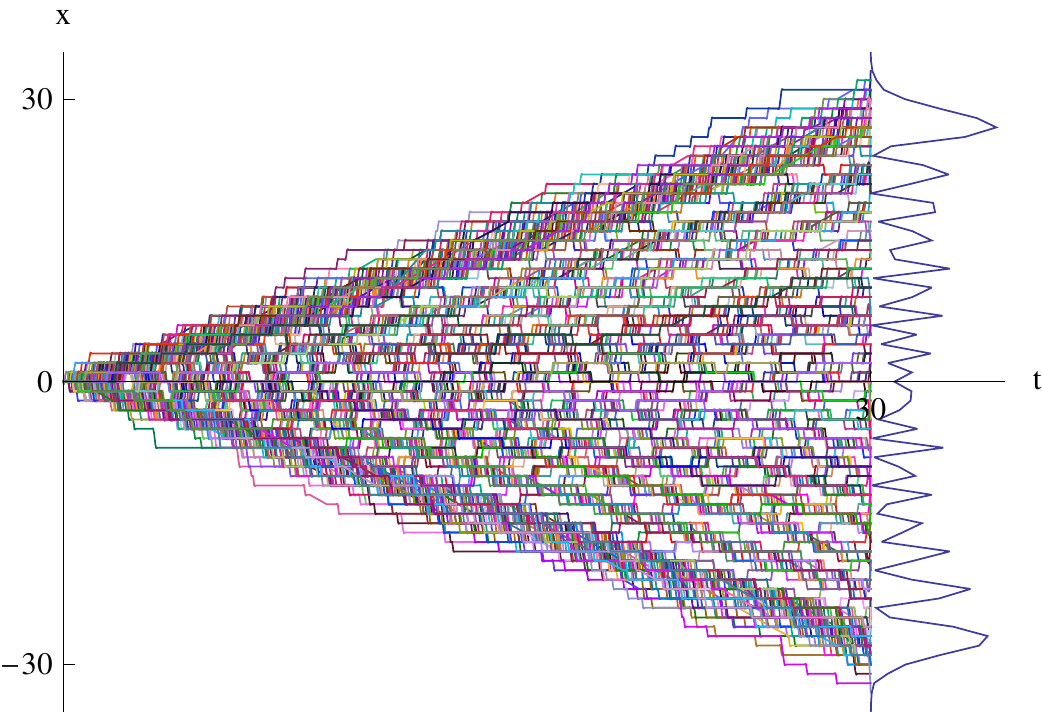}
\caption{A sample of $500$ paths of the stochastic process of \emph{Example 2}. For the purpose of comparison with figure \ref{fig:figura1}, the \emph{empirical distribution} at time $t=30$, of a sample of $5\cdot 10^4$ trajectories, is shown on the right.}
\end{figure}
\\[5pt]
The reader more interested in the \emph{phenomenon} than in the \emph{equations} (in this case the continuity equation \eref{eq:continuity} for the evolution of $\rho(t,x)$ and the forward Kolmogorov equation for the evolution of $p(t,x;t_0,x_0)$) will see, in section \ref{sec:autonomous}, how the numerical procedure leading to figure \ref{fig:samplepaths} actually makes use \emph{only} of a step by step implementation of the dynamical rules \eref{it:geometric}, \eref{it:flux}, \eref{it:attractor}.
\section{Quantum walks vs. random walks} \label{sec:quantum}

In this section we look at quantum mechanics as a metaphor suggesting, at the heuristic level, an interesting dynamical behaviour for a random (Markov) process exploring a graph or decision tree. We base our work on the classical results of Guerra and Morato  \cite{guerra83} on the formulation of quantum-mechanical behaviour in terms of controlled stochastic processes; the picture of a quantum walk that emerges through its stochastic analogue is that of a swarm of walkers moving according to transition rules involving the distribution of the \emph{entire} swarm.\\[5pt]
Consider a quantum system having as state space a Hilbert space the dimension of which we will indicate by $s$, and as generator of the time evolution a Hamiltonian operator that we will indicate by $H$.\\
Having fixed an orthonormal basis, $\ket{\phi_1}, \ket{\phi_2},\ldots, \ket{\phi_s}$, a graph $G$ is defined, \emph{starting from the selected basis and from the selected Hamiltonian}, by stating that $G$ has $\Lambda_s = \{1,. ... ,s\}$ as its set of vertices and edges $\{k, j\}$ such that $j \neq k$ and $|\bra{\phi_k} H \ket{\phi_j}| >0$.\\
In the context of this section, the graph $G$ will play the role played, in the more elementary context of \emph{Example 1} of section 1, by the linear graph having the relative integers as vertices, with edges between nearest neighbour sites. Similarly, the role played in section 1 by equation \eref{eq:discrschro} will be played in this section by the Schr\"odinger equation in the representation determined by the selected basis:
\begin{equation} \label{eq:newdiscrschro}
 i \frac{d}{dt} \psi(t,k) = \sum_{j=1}^s H_{k,j} \cdot \psi(t,j)
\end{equation}
with
\begin{equation}
 H_{k,j}=\bra{\phi_k} H \ket{\phi_j}.
\end{equation}
We pose in the following terms the question of finding in the general context of this section, an analogue of \emph{Example 2} of section 1:\\[5pt]
\emph{Easy problem}: find a constructive procedure associating with each solution $\psi$ of \eref{eq:newdiscrschro} a Markov process $q(t)$ on the graph $G$ having at each time $t$ probability distribution
\begin{equation}
 \rho(t,k)=P(q(t)=k)=|\psi(t,k)|^2.
\end{equation}
If this process exists and satisfies (for a suitable field $\nu$ of transition probabilities per unit time) the condition, that we impose as an analogue of conditions \eref{eq:semlambda} and \eref{eq:semmu},
\begin{equation}
 p(t+\tau,j;t,k)\equiv P(q(t+\tau)=j|q(t)=k) = \tau \cdot \nu_j(t,k)+o(\tau)
\end{equation}
for $\tau \to 0^+$ and for each $j$ being a neighbour of $k$ in the graph $G$, then it will satisfy the continuity equation
\begin{eqnarray} \label{eq:continuityrho} 
\frac{d}{dt}\rho(t,k) & = & \sum_{j \in N(k)} \rho(t,j) \nu_k(t,j)-\rho(t,k)\nu_j(t,k)=   \\
& = &  \sum_{j \in N(k)} (\rho(t,j) \nu_k(t,j)+ \rho(t,k)\nu_j(t,k)) \left (\frac{\rho(t,j) \nu_k(t,j)- \rho(t,k)\nu_j(t,k)}{\rho(t,j) \nu_k(t,j)+ \rho(t,k)\nu_j(t,k)} \right ). \nonumber
\end{eqnarray}
In the above equation, we have indicated by $N(k)$ the set of neighbours in $G$ of the vertex $k$, namely the collection of vertices $j$ such that $j \neq k$ and $\{j, k\}$ is an edge.\\
In the second line of equation \eref{eq:continuityrho} we have separated the term
\[
 \rho(t,j) \nu_k(t,j)+ \rho(t,k)\nu_j(t,k),
\]
symmetric in $j$ and $k$ (on the analogue of which we have imposed in Section 1 the \emph{geometric mean rule}), from the antisymmetric term 
\[
\frac{\rho(t,j) \nu_k(t,j)- \rho(t,k)\nu_j(t,k)}{\rho(t,j) \nu_k(t,j)+ \rho(t,k)\nu_j(t,k)},
\]
of absolute value $\leq 1$, representing the net relative flux of probability mass \emph{from $j$ into $k$}.\\
If, now, the same $\rho$ appearing in \eref{eq:continuityrho} satisfies also $\rho(t,k)=|\psi(t,k)|^2$  for a $\psi$ satisfying \eref{eq:newdiscrschro}, it must be
\begin{equation} \label{eq:debroglie}
 \psi(t,k)=\sqrt{\rho(t,x)} \exp(i \cdot S(t,k))
\end{equation}
for some phase function $S$ to be determined by inserting the Ansatz 
\eref{eq:debroglie} into equation \eref{eq:newdiscrschro}. Doing so, and separating the real and imaginary parts of the resulting equation, one gets two equations:
\begin{equation} \label{eq:eqesse}
 \frac{d}{dt} S(t,k) = -H_{k,k} -\sum_{j \in N(k)} h_{k,j}\sqrt{\frac{\rho(t,j)}{\rho(t,k)}} \cos(\beta_{k,j}(t))
\end{equation}
and
\begin{equation} \label{eq:eqrho}
 \frac{d}{dt} \rho(t,k) =\sum_{j \in N(k)} 2 h_{k,j}\sqrt{\rho(t,k)\rho(t,j)} \sin(\beta_{k,j}(t)),
\end{equation}
where we have set
\begin{equation}
 h_{k,j}=|H_{k,j}|
\end{equation}
and
\begin{equation}
 \beta_{k,j}(t)=Arg(H_{k,j})+S(t,j)-S(t,k).
\end{equation}
In order to check that our \emph{Easy problem} admits at least one solution, it is sufficient to compare the purely kinematic relations 
\begin{equation}
 \frac{d}{dt} \rho(t,k) =\sum_{j \in N(k)} 2 h_{k,j}\sqrt{\rho(t,k)\rho(t,j)} \sin(\beta_{k,j}(t)), \nonumber
\end{equation}
and
\begin{equation}
 \frac{d}{dt}\rho(t,k) =\sum_{j \in N(k)} (\rho(t,j) \nu_k(t,j)+ \rho(t,k)\nu_j(t,k)) \left (\frac{\rho(t,j) \nu_k(t,j)- \rho(t,k)\nu_j(t,k)}{\rho(t,j) \nu_k(t,j)+ \rho(t,k)\nu_j(t,k)} \right )\nonumber,
\end{equation}
viewed, for assigned $\psi$ and therefore for assigned $\rho$ and $\beta$, as constraints on the unknown transition probabilities per unit time of the process $q(t)$ to be constructed. The simplest way to satisfy this constraint is by requiring term by term equality in the sums that appear in the right hand sides, and by equating in each term the symmetric and antisymmetric factors.\\
We thus get the equations
\begin{eqnarray}
 \rho(t,j) \nu_k(t,j)+ \rho(t,k)\nu_j(t,k) & = & 2 h_{k,j}\sqrt{\rho(t,k) \rho(t,j)}  \label{eq:eqsymm}\\
 \rho(t,j) \nu_k(t,j)-\rho(t,k)\nu_j(t,k) & = & 2 h_{k,j}\sqrt{\rho(t,k) \rho(t,j)} \sin(\beta_{k,j}(t))\label{eq:eqantisymm}
\end{eqnarray}
that are solved by
\begin{eqnarray} \label{eq:hh}
 \nu_k(t,j) & = & h_{k,j}\sqrt{\frac{\rho(t,k)}{\rho(t,j)}} \left( 1+\sin(\beta_{k,j}(t))\right) = \nonumber \\
& = & h_{k,j}\sqrt{\frac{\rho(t,k)}{\rho(t,j)}} \left( 1+\sin(Arg(H_{k,j})+S(t,j)-S(t,k))\right)
\end{eqnarray}
for $k \in N(j)$.\\
For more details, and for the physical motivation (related to questions of time reversal invariance) of the merits of this particular choice, we refer  to \cite{guerra84}.\\
It is immediate to check that \eref{eq:eqsymm} is precisely the \emph{geometric mean rule} \eref{it:geometric} of section 1.\\
It is also an easy exercise to check that \eref{eq:hh} specializes, due to the phase factor $i^x$ in equation \eref{eq:solschro}, to conditions \eref{it:flux} and \eref{it:attractor} in the simple context of section 1.
\section{Autonomous generation} \label{sec:autonomous}
\begin{figure}[!h]
	\centering
		\subfigure[]{\label{fig:figura3a} \includegraphics[width=5cm]{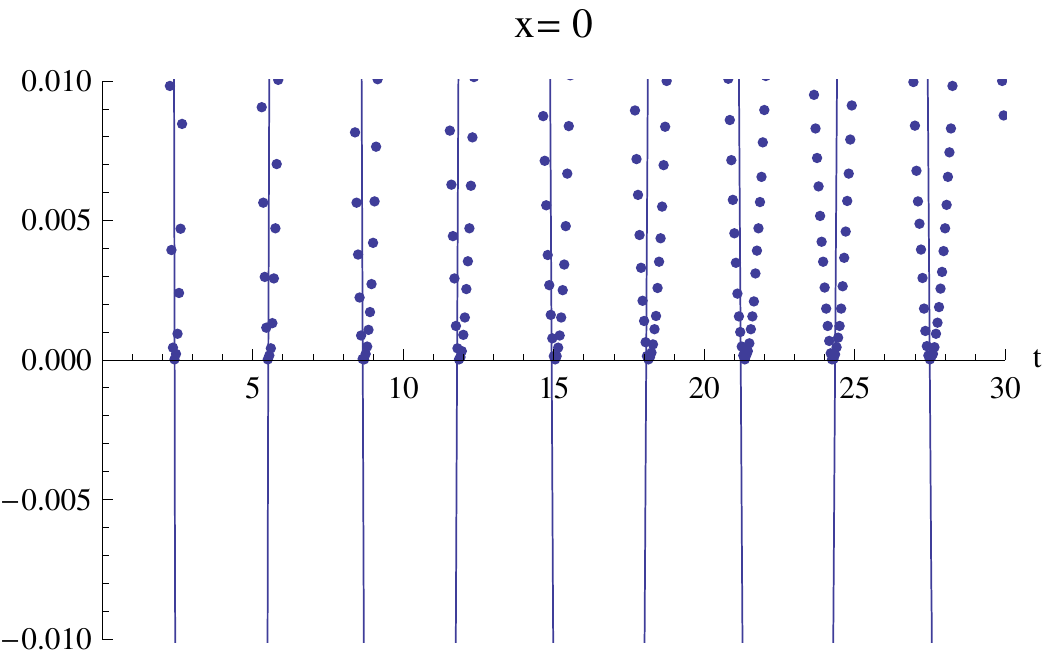}}
		\hspace{0.5cm} 
		\subfigure[]{\label{fig:figura3b} \includegraphics[width=5cm]{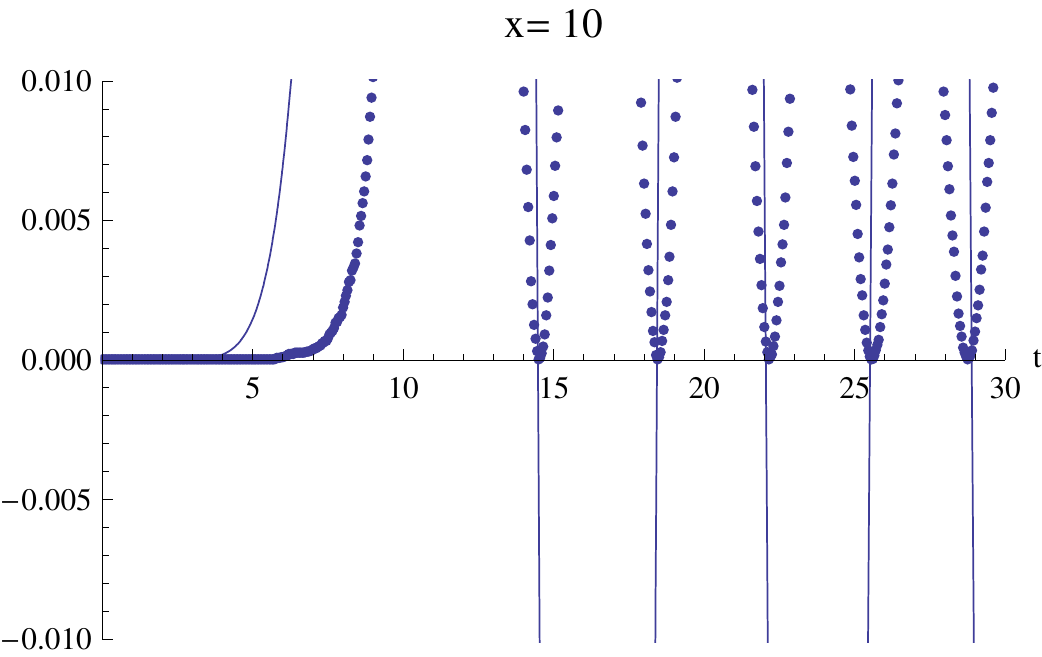}}
		\subfigure[]{\label{fig:figura3c} \includegraphics[width=5cm]{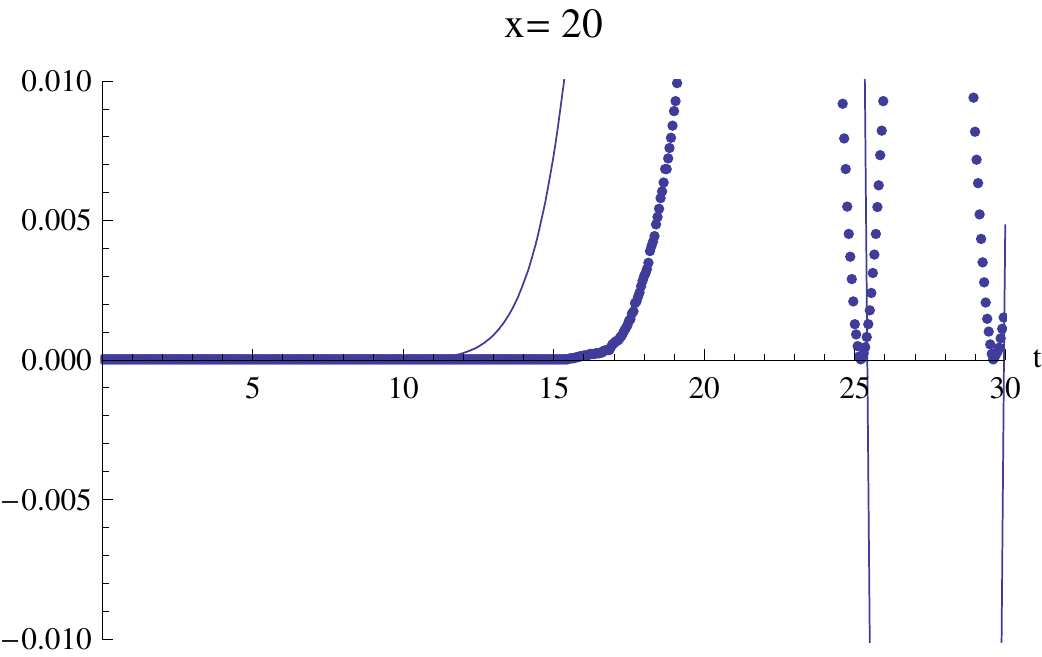}} 
		\hspace{0.5cm}
		\subfigure[]{\label{fig:figura3d} \includegraphics[width=5cm]{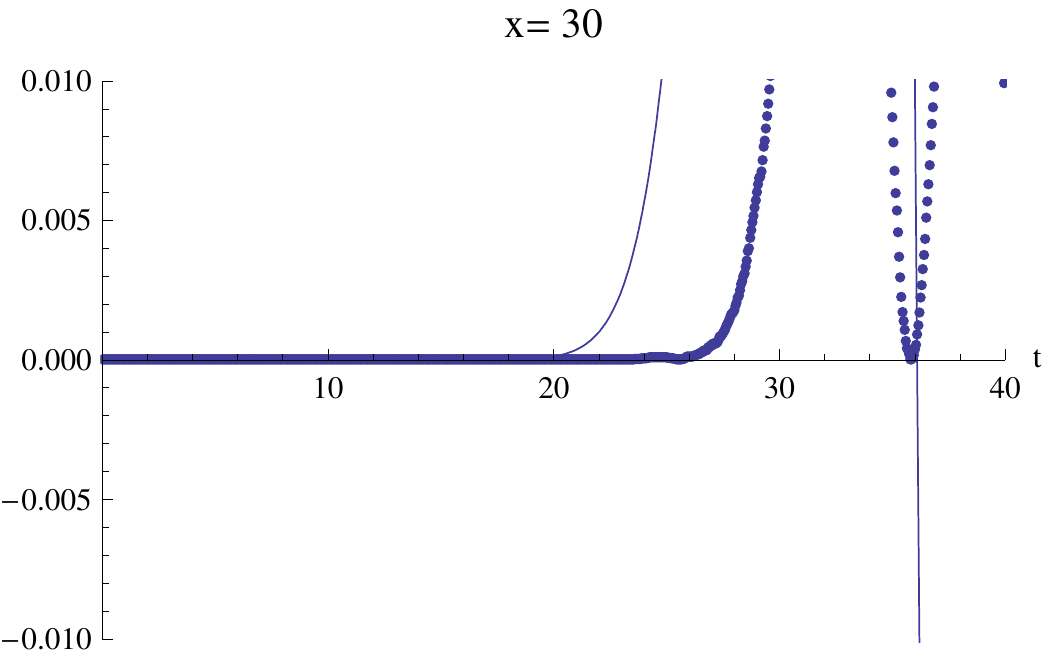}}
\caption{Thin solid lines: graphs of $J_x(t)$ as a function of $t$ for several values of $x$. Thick dashed lines: graphs, as a function of $t$, of the fraction of trajectories that visit $x$ at time $t$.}
	\label{fig:figura3}
\end{figure}

The \emph{Hard problems}, as opposed to the kinematical \emph{Easy problem} reviewed in section \ref{sec:quantum}, are
\begin{enumerate}
\def\theenumi{\Roman{enumi}}
\item \label{it:dynamical} understand \eref{eq:eqesse} as a dynamical condition on the processes $q(t)$ that solve our \emph{Easy problem};
\item \label{it:auto} \emph{autonomously} simulate these processes by actual implementation of this dynamical condition.
\end{enumerate}
Problem \ref{it:dynamical} is discussed in full detail in  \cite{guerra84} following the general approach of \cite{guerra83} in which stochastic control theory is successfully proposed as a very simple model simulating quantum-mechanical behaviour.\\
We are not able to tackle problem \eref{it:auto} in its generality. We can only go back to section 1  and show that the three dynamical rules and the initial conditions stated there in assigning \emph{Example 2} are enough to generate the sample paths of figure \ref{fig:samplepaths}.\\
This is far from obvious because of the \emph{geometric mean rule}: it requires, in the numerical generation of each sample path, to take into account the \emph{causative effect} (through the \emph{estimated} probability distribution) on each trajectory of the \emph{ensemble} of trajectories to which it belongs \cite{smolin06} .\\
Even to show, as in figure \ref{fig:samplepaths}, a small sample of trajectories, the need of carefully estimating at each time step the density $\rho$ imposes the simultaneous generation of a large number $N_{tr.}$ of trajectories.\\
The numerical procedure leading to figure \ref{fig:samplepaths}, makes, by purpose, no reference to the solution of the continuity equation we have given in section 1, nor to the solution of the Kolmogorov equations for the conditional probabilities \mbox{ $p(t,x;t_0,x_0)$} that can be easily found by similar techniques. We present here this procedure in some detail because the challenges one meets in simulating the process $q(t)$ by implementing rules \eref{it:geometric}, \eref{it:flux}, \eref{it:attractor} and the initial conditions listed in section 1 give an operational meaning to the notion of \emph{autonomous} simulation.\\[5pt]
The state of the system at each time $t=\tau \cdot k$, where the integer $k$ runs from $1$ to $n_{steps}$ and $\tau$ is the time step, is described by the pair
\begin{itemize}
\item \emph{configuration array} of length $N_{tr}$: its $j$-th element $q_j(t)$ indicates the current position of the $j$-th trajectory; a space cut-off is introduced through an integer parameter $L$ such that each trajectory is followed as long as \mbox{$-L\leq q(t) \leq L$}; the empirical density  $\rho_{emp}$ of the process at each time is estimated from the \emph{configuration array};
\item \emph{transition array} indexed from $-L$ to $L$: its $x$-th element is an ordered pair of bits $(m_x,l_x)$: if $m_x=1$ (resp. $l_x=1$) then transitions $x \to x-1$ (resp. $x \to x+1$) are allowed, whereas if $m_x=0$ (resp. $l_x=0$) they are forbidden.
\end{itemize}
In our implementation $n_{tr.}=5 \cdot 10^4$, $\tau=0.05$, and the process has been followed up to time $t_{max}=\tau \cdot n_{steps}=100$, well beyond the time window shown in figure \ref{fig:samplepaths}; the space cut-off has been set at $L = 150$.\\
The algorithm consists of the iteration $n_{steps}$ times of the following steps:
\begin{enumerate}
\item estimate $\rho_{emp}$ from the \emph{configuration array};
\item increment each $q_j(t)$ by $Move(t,q_j(t))$, where the random variable $Move(t,x)$ takes the values $-1,0,+1$ with probabilities $\tau \cdot \mu_{emp}(t,x),1-\tau \cdot (\mu_{emp}(t,x)+\lambda_{emp}(t,x)), \tau \cdot \lambda_{emp}(t,x)$, respectively.\\
The empirical transition rates $\lambda_{emp}$ and $\mu_{emp}$ are here given by
\begin{subeqnarray} \label{eq:stime}
\lambda_{emp}(t,x) & = & \sqrt{\frac{\rho_{emp}(t,x+1)}{\rho_{emp}(t,x)}}\ l_x; \slabel{eq:stimal}\\
\mu_{emp}(t,x)& = &  \sqrt{\frac{\rho_{emp}(t,x-1)}{\rho_{emp}(t,x)}}\ m_x \slabel{eq:stimam},
\end{subeqnarray}
\item estimate the new empirical distribution $\rho_{emp}(t+\tau,x)$;
\item if $\rho_{emp}(t,x)>0$ and $\rho_{emp}(t+\tau,x)=0$ then update the \emph{transition array} following the \emph{``horror vacui'' rule} \eref{it:attractor}, namely setting $l_{x-1}=1$ and $m_{x+1}=1$, and restore the \emph{local unidirectionality rule} by setting $(m_x,l_x)=(0,0)$.
\end{enumerate}
In the initialization step the \emph{transition array} has been given the initial assignment
\begin{eqnarray}
(m_x,l_x)& = &\begin{cases}
		(1,0) \mbox{ if $x<0$} \\
		(1,1) \mbox{ if $x=0$} \\
		(0,1) \mbox{ if $x>0$}
          \end{cases} \nonumber
\end{eqnarray}
Before discussing the initialization of the \emph{configuration array}, we observe that the rough first order updating rule of step 2. runs into trouble if a proposed transition involves a site at which $\rho_{emp}$ vanishes, the problem being with zeroes of the \underline{numerators} under the square root of \eref{eq:stimal} and \eref{eq:stimam}. The most evident form of this fact is that, given that the process at time $0$ is at position $0$, the probability that it moves at all in a time step is
\begin{equation}
1-J_0^2(\tau)=\frac{\tau^2}{2}+O(\tau^4).
\end{equation}
We have found an inexpensive way out of this difficulty by initializing the \emph{configuration array} by the assignment:
\begin{eqnarray}
q_j(0) & = & 0, \mbox{ for } j=(2L+1)+1 \ldots, N_{tr}. \nonumber\\
q_j(0) & = & j-L-1, \mbox{ for } j=1,\ldots, 2L+1  \nonumber
\end{eqnarray}
The first line says that most of the trajectories start from the origin; the second that trajectory 1 starts from $-L,\ \ldots$, trajectory $2L+1$ starts from $L$.\\
The second line makes sure that initially there is at least one trajectory per site; this situation is restored, after step 4. by:
\begin{enumerate}
\setcounter{enumi}{4}
\item set $q_j(t+\tau)= j-L-1, \mbox{ for } j=1,\ldots,2L+1$.
\end{enumerate}
The dummy trajectories labelled by $j=1,\ldots,2L+1$ provide some probability mass when needed to prevent the first order procedure from getting stuck.\\
Comparison between figures \ref{fig:figura1} and \ref{fig:samplepaths} gives an idea of how well our simple procedure fills the \emph{configuration array} .\\
An analogous comparison is conducted in figure \ref{fig:figura3} for the \emph{transition array}: the issue there is how well our procedure catches the instants of time at which the \emph{control mechanism} expressed by the ``\emph{horror vacui'' rule} takes hold, namely the zeroes of $J_x(t)$.
\section{Conclusions and outlook}
If you give me a quantum walk efficiently exploring a graph or decision tree \cite{farhi97}, I take your computational basis, your initial condition and your Hamiltonian and cook for you a stochastic process by computing its transition probabilities per unit time according to the recipe of section 2 and its transition probabilities  $p(t,x;t_0,x_0)$ by integration of the Kolmogorov equations (this can be done in quite explicit terms for the \emph{Example 2} of section 1) or by a clever exploitation of a few rules controlling the dynamics, as done in section 3. The discussion of section 2 makes it clear that my \emph{random} walk will, by construction, visit your graph or decision tree as efficiently as your \emph{quantum} walk.\\[5pt]
Can  the above statement be reconciled with the statement that the quantum \emph{glued trees} algorithm of  \cite{farhi03}  outperforms any classical algorithm? How are the \emph{classical alternatives}  defined in the original literature on exponential speedup by quantum walk?  Does the \emph{causative effect of the ensemble} disqualify a Markov process from being classical ?\\[5pt]
On these points, all we can do is to advance a conjecture: the cost of my random simulation of your quantum walk is hidden in the size $N_{tr.}$ of the sample I am required to generate. We have indeed called attention, since section 1, on the \emph{geometric mean rule}: for every edge of your graph the probability per unit time of a transition of my process along that edge is equal to the geometric mean of the probabilities of the process at the two vertices joined by that edge.\\[5pt]
We pose as a problem of future research the quantitative assessment of the cost (as measured by  $N_{tr.}$) of the density estimation step required before each updating in the simulation.

\bibliography{mybib}
\bibliographystyle{unsrt}
\end{document}